\documentstyle[12pt]{article}
\newtheorem{theorem}{Theorem}
\newtheorem{lemma}[theorem]{Lemma}
\newtheorem{corollary}[theorem]{Corollary}

\newtheorem{example}[theorem]{Example}

\newenvironment{proof}{{\bf Proof:\ \ }}{\hfill$\bull$\medskip}

\newfont{\gothic}{eufm10 scaled\magstep1}

\newcommand{\A}{\alpha}
\newcommand{\B}{\beta}
\newcommand{\C}{\gamma}
\newcommand{\D}{\delta}
\newcommand{\E}{\epsilon}

\newcommand{\LA}{\lambda}

\newcommand{\R}{\rho}

\newcommand{\complex}{{\bf C}}
\newcommand{\dd}{\partial}

\newcommand{\gR}{{\gothic R}}

\newcommand{\half}{{1\over 2}}

\def\sqr#1#2{{\vcenter{\vbox{\hrule height.#2pt 
        \hbox{\vrule width.#2pt height#1pt \kern#1pt
           \vrule width.#2pt}                   
        \hrule height.#2pt}}}}                  
\def\tomb{\sqr84}

\newcommand{\rf}[1]{(\ref{#1})}

\newcommand{\beq}{\begin{equation}}
\newcommand{\bex}{\begin{example}\rm}
\newcommand{\barray}{\begin{eqnarray}}
\newcommand{\eeq}{\end{equation}}
\newcommand{\eex}{\end{example}}
\newcommand{\earray}{\end{eqnarray}}
\newcommand{\nn}{\nonumber}

\begin{document}
\begin{center}
{\large\bf
THE COALESCENCE LIMIT OF THE 
SECOND PAINLEV\'E EQUATION}
\end{center}
\centerline{{\bf Rod Halburd} \hskip 2mm and \hskip 2mm  {\bf Nalini Joshi}}
{\small\begin{center}
School of Mathematics,
University of New South Wales\\
Sydney NSW
Australia 2052
\vskip 2mm
{\small
e-mail: rodney@solution.maths.unsw.edu.au or N.Joshi@unsw.edu.au
\vskip 2mm
AMS classification numbers: 34A12, 34A20, 34A25, 34E10
}
\end{center}}
\vskip 5mm
\begin{abstract}
In this paper, we study a well known asymptotic limit in which the
second Painlev\'e equation ($P_{II}\/$) becomes the first Painlev\'e
equation ($P_I\/$).  The limit preserves the Painlev\'e property
(i.e. that all movable singularities of all solutions are poles).
Indeed it has been commonly accepted that the movable simple poles
of opposite residue of the generic solution of $P_{II}\/$ must
coalesce in the limit to become movable double poles of the
solutions of $P_I\/$, even though the limit naively carried out on the
Laurent expansion of any solution of $P_{II}\/$ makes no sense.
Here we show rigorously that a coalescence of poles occurs.  Moreover
we show that locally all analytic solutions of $P_I\/$ arise as limits
of solutions of $P_{II}\/$.
\end{abstract}
\section{Introduction}
An ordinary differential equation is said to be of Painlev\'e type
(or to possess the Painlev\'e property) if the only movable singularities
of its solutions are poles.  The property
is strongly related to integrable systems (systems which can be solved
via related linear problems)
\cite{ablowitzone, ablowitztwo, kruskalclarkson, review}.
Knowledge of equations with the Painlev\'e property, including various
methods of classification, is therefore valuable in the search for
integrable systems.  Asymptotic limits of differential equations that
preserve the Painlev\'e property provide another mechanism for such
searches.

In a series of papers published around the turn of the century,
Painlev\'e \cite{painlevestuff}, Gambier \cite{gambier}, and Fuchs
\cite{fuchs}
conducted an exhaustive search for all equations of Painlev\'e type of the
form
$$	u''=\Phi(x;u,u'),	$$
where $\Phi$ is analytic in $x$ and rational in $u$ and $u'\/$.
They discovered six equations of Painlev\'e type which, up to a
transformation,
are the only equations of this type whose general solutions are new
transcendental functions.
These equations are known as the
Painlev\'e equations $P_I$--$P_{\,V\!I}$.  The first two are
$$
\begin{array}{ccccrclccccc}
        P_{{I}}: &\qquad&\qquad& \qquad& u''(x) &=& 6u^2+x; &\qquad &
\qquad & \qquad &\qquad &\qquad\\
        P_{{II}}: &\qquad&\qquad &\qquad& u''(x) &=& 2u^3+xu+\A; &\qquad &
\qquad &\qquad &\qquad &\qquad
\end{array}
$$
where $\A$ is a complex constant.

Painlev\'e \cite{painleve} noted that under the transformation
\barray
        x &=& \E^2 z-6\E^{-10};\nn\\
        u &=& \E y+\E^{-5};\label{polestrans}\\
        \A &=& 4\E^{-15},\nn
\earray
$P_{II}$ becomes
\beq    y''(z) = 6y^2+z+\E^6\left\{2y^3+zy\right\}. \label{polespiie}\eeq

If $\E$ vanishes, equation \rf{polespiie}, which we will refer to as
$P_{II}(\E)\/$, becomes $P_I$
with $x$ replaced by $z$ and $u$ replaced by $y$.
We will say that under \rf{polestrans}, $P_{II}$ degenerates to $P_I$ and
write
$P_{II} \rightarrow P_I$.  Painlev\'e gave a series of such degeneracies which
is summarized in Figure~1.

  \begin{figure}[h]
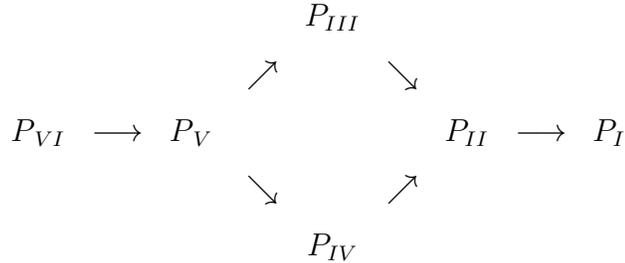

  $$
  \begin{array}{ccccccccc}
  & &     &       & P_{III}       &       &       & & \\
  & &     &\nearrow &     &\searrow       &       & & \\
  P_{\,VI} & \longrightarrow & P_{\,V}
  \hskip 1mm &    &       &       &P_{II} &
  \longrightarrow & P_I\\
  & &       &\searrow &     &\nearrow       &       & & \\
  & &       &       & P_{IV}       &       &       & & \\
  \end{array}
  $$
  \caption{Degeneracies among the Painlev\'e equations.}
  \end{figure}
 Transformations which lead to these degeneracies can be motivated
 from the point of view of isomonodromy problems \cite{garnier, okamoto}.
Using a method based on maximal dominant balances, Joshi and Kruskal
\cite{joshi} have found a new degeneracy of $P_{IV}$ to
another equation of Painlev\'e type (equation XXXIV on p.340 of Ince 
\cite{ince}).  Their paper raises the possibility of using
asymptotic limits
between differential equations which preserve the Painlev\'e property
as new tools in the search for, and classification of, integrable systems.

The central concern of the present paper is an exploration of the convergence
of solutions of $P_{II}(\E)$ to solutions of $P_I\/$ as $\E$ vanishes.
In particular
we are concerned with the way in which simple
poles of oppositely signed residues in solutions of $P_{II}(\E)$ coalesce to
form the double poles of $P_I\/$. 
Unfortunately, a purely local analysis of this coalescence is problematic as
the radius of convergence of any Laurent expansion centred on any pole 
necessarily decreases to zero if the poles coalesce.  Rather than attempt to
find an accurate upper bound on the radius of convergence
we use techniques based on steepest ascent curves similar to those
first expounded by Joshi and Kruskal in their direct proof that $P_I$ to
$P_{\,VI}$ possess the Painlev\'e property \cite{joshidirect}.  Before
embarking
on this problem we analyse a model problem given by a similar degeneracy
between the
autonomous
versions of $P_I$ and $P_{II}$ whose general solutions are expressible
in terms of elliptic functions.
We show that the poles here coalesce by estimating
the distance between them.  Such estimates are
obtained in section 2.

In section 3 we consider general equations of
the form
\beq    {dy_i\over dz}=f_i(z,y_1,\ldots,y_n;\E),\qquad 1\le i \le n
\label{polesdiff}\eeq
where the $f_i$ are entire functions of $(z,y_1,\ldots,y_n;\E)$.  We
will say that equations \rf{polesdiff} degenerate to the equations
\beq    {dy_i\over dz}=f_i(z,y_1,\ldots,y_n;0),\qquad 1\le i \le n
\label{polestarget}\eeq
in the limit as $\E$ approaches zero.  We show that, locally, any
analytic solution of the target equations \rf{polestarget} can be obtained
in the limit as $\E\to 0$ of a solution to equations \rf{polesdiff}.
A corollary of the theorem states that if $y_I$ is a solution of $P_I$ then
given any compact subset $K$ on which $y_I$
is analytic then there is a solution,
$y$ of $P_{II}(\E)$ such that $y\to y_I$ on $K$ with respect to the
sup norm as $\E\to 0\/$.
Hence, by considering the maximal analytic extension of $y$ we see that
$y\to y_I$ everywhere.
In section 4 we examine the rate of coalescence of
poles.  We obtain estimates of the distances between coalescing
poles and show that these are of order $\E^3\/$.

\section{Two Autonomous Painlev\'e Equations}

Consider the following autonomous versions of $P_I$ and $P_{II}$,
\barray
	E_I\qquad\qquad\qquad\qquad\qquad
	& &
	u'' = 6u^2+\LA\qquad\qquad\qquad\qquad\qquad\qquad\qquad
\qquad\qquad\qquad\nn\\
	E_{II}\qquad\qquad\qquad\qquad\qquad
        & &
	u'' = 2u^3+\mu u+\A\qquad\qquad\qquad\qquad\qquad\qquad\qquad
\qquad\qquad\nn
\earray
where $\LA,\mu\in\complex$ are constants and the primes denote differentiation
with respect to $x\/$.  The solutions of $E_I$ and $E_{II}$
are either constants or may be expressed in terms of elliptic integrals.

Following the analogy of the $P_{II}\to P_{I}$ coalescence we transform the
variables in $E_{II}$ as follows:
$$	x=\E^2z,\quad \mu=\lambda\E^2-6\E^{-10},\quad u=\E y+\E^{-5},
\qquad\A=4\E^{-15}.$$
Under this transformation $E_{II}$ becomes
\beq	\ddot y=6y^2+\lambda+\E^6\left( 2y^3+\lambda y\right),
\label{papelliptic}\eeq
where a dot denotes differentiation with respect to $z\/$,
giving us the degeneracy $E_{II}\to E_I$.  In order to examine the nonconstant
solutions of \rf{papelliptic} we multiply the equation through by $\dot y$
and integrate.  In this way we obtain
\beq	{\dot y}^2=\E^6P_\E(y):=h+2\lambda y+\E^6\lambda y^2+4y^3+
\E^6y^4 \label{papint}\eeq
where $h\in\complex$ is a constant of integration.
Take $h$ given and fixed in the following analysis.
The nonconstant
solutions of equation \rf{papint} satisfy
$$	\E^3{dz\over dy} = Q_\E(y):=\frac1{\sqrt{P_\E(y)}}.$$
Now, for $\E\ne 0\/$,
$$      P_\E(y)=:(y-a_0)(y-a_1)(y-a_2)(y-a_3),   $$
where
\barray
a_0 =a_0(\E)&=&	-\frac4{\E^6}+\frac\lambda8\E^6+O(\E^{12}),
\label{papaexpan}\\
a_i =a_i(\E)&=& \eta_i+O(\E^6),\qquad i=1,2,3\nn
\earray
are the zeros of $P_\E(y)$ and the $\eta_i$ are zeros of $P_0(\eta)\/$.

We briefly recall some of the standard results from the theory
of elliptic integrals, beginning with a description of a Riemann surface for
$Q_\E(y)$ (see, for example Siegel \cite{siegel}).
We will assume that $h$ is such that for small $\E$, $P_\E(y)$
has distinct zeros (note that this is the generic case).
Cut two nonintersecting slits in the Riemann sphere,
say one from $a_0$ to $a_1$ and the other from $a_2$ to $a_3\/$. Make two
copies of the resulting manifold and label them $\mbox{\gothic M}_1$ and
$\mbox{\gothic M}_2$; these two
slit spheres correspond to the two branches of the square root operation
in the definition of $Q_\E(y)\/$.  Now take each side of both slits on
$\mbox{\gothic M}_1$ and identify them with the opposite sides of the
corresponding slits of $\mbox{\gothic M}_2\/$.  The resulting Riemann surface,
\gR, is homeomorphic to the 2-torus $T^2$. 
$Q_\E(y)$ is meromorphic throughout \gR\ and the elliptic integral
$$	I(\C):=\int_\C Q_\E(y)dy$$
is well defined for any piecewise smooth curve $\C$ in \gR\ where $\tilde y$
varies over the natural projection of $\C$ to the Riemann sphere ${\bf
CP}^1\/$.

Suppose that $y$ has poles with residues of opposite sign at $z_+$ and
$z_-\/$, then
\beq z_+-z_-=\E^{-3}\int_{\C}Q_\E({y}) dy,
\label{papopppoles}\eeq
where $\C$ is a path connecting $\infty_1$ and $\infty_2\/$ --- the
subscripts distinguish the points
at infinity on the two slit spheres $\mbox{\gothic M}_1$ and
$\mbox{\gothic M}_2$ respectively.  
Such a path must pass through one of the open slits
connecting the two spheres.  Its projection onto the
Riemann sphere must loop around the points $a_k\/$, $k=0,\ldots,3\/$
an odd number
of times
(note that if it encloses an even number of the
points $a_k\/$, the resultant integral is just a period of the elliptic
function $y\/$).  For small $\E$ the point $a_0$ is closest to
infinity so we consider a path which begins at $\infty_1$ and remains
in $\mbox{\gothic M}_1$ until it reaches the point $a_0\/$, loops around it,
and then retraces
the corresponding path in $\mbox{\gothic M}_2\/$, terminating at $\infty_2\/$.
Since an arbitrarily small loop around $a_0\/$ contributes nothing to
\rf{papopppoles}, the distance between the two poles is simply
\beq	|z_+-z_-|=2\left|\E^{-3}\int_{\infty}^{a_0}Q_\E({y}) dy
\right|.\label{papdist}\eeq

Next we refine our choice of the
path of integration for the right hand side of equation~
\rf{papdist}.  
At any point where $y$ is analytic and neither $y$ nor $y'$
vanishes, there is a unique direction of fastest increase in $|y|\/$.
Hence we can define a steepest ascent curve through any such point.
A simple calculation using the Cauchy-Riemann equations shows that on such
a curve, $d|y|=|dy|\/$ (on a path of steepest descent, $d|y|=-|dy|\/$).

Let $R_+$ be the connected component of the region
$$\Omega:=\{z: \ 
|y(z)|>|a_0|\}$$
 containing $z_+$ in its closure.  Note that
the only pole in the closure of $R_+$ is $z_+\/$ (since there is a unique
level curve of $|y|$ passing through every point in $R_+\/$).  Expanding
$y$ about a point $z_1$ in the boundary of $R_+$ such that $y(z_1)=a_0\/$
gives
$$      y(z)=a_0+\half y''(z_1)(z-z_1)^2+O\left( (z-z_1)^3\right),      $$
where $a_0\approx -4\E^{-6}$ and, from equation \rf{papelliptic},
$y''(z_1)\approx -32\E^{-6}\/$.  We see that $z_1$ is a (complex)
saddle point.
This implies that $z_1$ is the initial point for two steepest ascent
curves (and two steepest descent curves).  One of the steepest ascent
curves must enter $R_+$ and terminate at $z_+\/$.  This is the path,
$\Gamma\/$, over which we integrate in equation \rf{papdist}.

Choose $r>0$ so small that for all $\E$ such that $|\E|<r\/$,
$$	|a_0|\ge2\max_{1\le i\le 3}\left\{|a_i|\right\}.	$$
Note that this can be achieved because the
expansion \rf{papaexpan} shows that $a_0$ is large for small $\E\/$.
Then, since $|y(z)|>|a_0|$ on $\Gamma\/$,
for $|\E|<r\/$ and $1\le i\le 3\/$, we get
$$ |y|\le |y-a_i|+|a_i|\le |y-a_i|+|y|/2\quad\Rightarrow\quad
|y|/2\le |y-a_i|.$$
So
\beq	\left|Q_\E(y)\right|\le
	{2\sqrt2\over\sqrt{|y|^3\left(|y|-|a_0|\right)}}\label{papQ}.\eeq
In particular, notice that the integral in \rf{papdist} is convergent at
infinity.

%
Using  \rf{papQ}, we find
from equation \rf{papdist} that 
\barray
	|z_+-z_-| &\le& -2|\E|^{-3}\int_\infty^{|a_0|}
	{2\sqrt2 \,\,d\!\left|y\right|\over
	\sqrt{|y|^3\left(|y|-|a_0|\right)}}\nn\\
	&=& -{8\sqrt2\over |a_0|| |\E|^3}\sqrt{1-{|a_0|\over |y|}}
\Bigg|^{y=a_0}_{y=\infty}
	={8\sqrt2\over |a_0(\E)| |\E|^3}=O(\E^3).\nn
\earray
Therefore the two oppositely signed poles
coalesce as $\E$ vanishes.

In the above analysis we have only considered the generic case in which
$P_\E(y)$ has four distinct zeros.  In the nongeneric case the 
Riemann surface, \gR, of $y$ is no longer a torus.  Our analysis, however,
does not depend critically on the global topology of \gR{} and the same
estimates apply.

\section{Local Analytic Solutions}
The aim of this section is to prove the following theorem:-
\begin{theorem}
Let $(\eta_1\ldots,\eta_n)$ be a given solution of the system of ODEs in
\rf{polestarget} which is analytic in some pathwise connected region
$\Omega\subseteq\complex$ and choose $z_0\in\Omega$.  
Given any simply
connected
compact subspace $K\subset\Omega$ containing $z_0$,
there exists a solution
$(y_1,\ldots,y_n)$ of equations \rf{polesdiff} and a number 
\hskip 1mm $r_K>0$ such that,
\begin{enumerate}
\item the $y_i$ are analytic in $(z,\E)$ for
$z\in K\/$, $|\E|<r_K\/$;
\item $y_i(z,0)=\eta_i(z)$ $\forall z\in K\/$;
\item $y_i(z_0,\E)=\eta_i(z_0)$ $\forall\E$ such that $|\E|<r_K\/$.
\end{enumerate}
\end{theorem}
Note that, regardless of the choice of $K\/$, the $y_i$ satisfy the same
initial value problem at $z_0\/$.
This theorem shows us that, locally, solutions of equations \rf{polesdiff}
converge onto solutions of equations \rf{polestarget}.  It shows that the
singularities of this family of solutions of equations \rf{polesdiff}
lie arbitrarily close to those of equations \rf{polestarget} (or go to
infinity), for small $\E\/$.  In the proof of this theorem given
below we will make use of
the following lemma which can be proved using elementary arguments involving
majorant series (see, for example, Cartan \cite{cartan}).

\begin{lemma}
Consider the system of ODEs
\beq    {dy_i\over dz}=f_i(z,y_1,\ldots,y_n;\E),\qquad 1\le i \le n
\label{polesdiffe}\eeq
together with the initial conditions
$$      y_i(z_0,\E)=\phi_i(\E),\qquad 1\le i \le n      $$
where the $\phi_i$ are analytic for $|\E|\le r$ and the $f_i$ are
analytic on
$$      S:=\{(z,y_1,\ldots,y_n;\E): |z-z_0|\le\R,\ |y_i-\phi_i|\le R,\
|\E|\le r,\ 1\le i\le n\}.$$
Then there is a unique solution ${\bf y}:=(y_1,\ldots,y_n)$ of
\rf{polesdiffe} which is analytic in $(z,\E)$ whenever $|\E|<r$ and
$$ |z-z_0|<Z_{\R,r,R}^M(\E):=\R\left(1-\exp \left[ -{(1-|\E|/r)R\over
        (n+1)\R M}\right]\right),       $$
where
$$      M\ge \sup_S |f_i|,\qquad 1\le i\le n.      $$
\label{polesexist}
\end{lemma}

\noindent{\bf Proof of Theorem 1:}
Since the $f_i$ are entire, we may expand them as power series,
$$      f_i(z,y_1,\ldots,y_n;\E)=
        \sum a^i_{j k_1\cdots k_n l}z^jy_1^{k_1}\cdots y_n^{k_n}\E^l,   $$
which converge everywhere.

Fix $\R,R,r_0>0$.  Let $\Gamma$ be any finite length
curve connecting $z_0$ to $\dd K$.  We will first prove existence in a
thin neighbourhood of $\Gamma$.  Define
\barray
        B &:=&  \sup\{ |z| : |z-\tilde z|=\R,\ \tilde z\in\Gamma\},\nn\\
        L &:=&  {\sup_{\scriptstyle z\in \Gamma\atop
1\le i\le n}}|\eta_i(z)|,\nn
\earray
and $M:=\max_{1\le i\le n}M_i$ where
\barray
        M_i &:=& 2\sum |a^i_{jk_1\cdots k_n l}|B^j(R+L)^{k_1\cdots k_n}r_0^l.
\earray
This last series converges because the $f_i$ are entire.

Let $S_0=\{(z,y_1,\ldots,y_n;\E): |z-z_0|\le\R,\ |\E|\le r_0,\ |y_i-
\eta_i(z_0)|\le R,\ 1\le i\le n\}$.  Then for $(z,y_1,\ldots, y_n;\E)\in
S_0$ we have
\barray
& &     |f_i(z,y_1,\ldots y_n;\E)| \nn\\
&\le&   \sum    |a^i_{jk_1,\cdots k_nl}| |z|^j
(|y_1-\eta_1(z_0)|+|\eta_1(z_0)|)^{k_1}\!\cdots
(|y_n-\eta_n(z_0)|+|\eta_n(z_0)|)^{k_n} |\E|^l\nn\\
&\le&   \sum    |a^i_{jk_1,\dots k_nl}| B^j(R+L)^{k_1\cdots k_n}r_0^l\nn\\
&=& \half M_i.\nn
\earray
Therefore $\sup_{z\in S_0}|f_i|\le M$ and so we deduce from Lemma
\ref{polesexist} that
there is a solution ${\bf y}^{(0)}:=(y^{(0)}_1,\ldots,y^{(0)}_n)$
of equations \rf{polesdiff}
satisfying the initial condition
$$      {\bf y}^{(0)}(z_0,\E)={\bf\eta}(z_0),\qquad |\E|<r_0.$$
Furthermore, ${\bf y}^{(0)}$ is analytic in $(z,\E)$ provided
$|z-z_0|<Z_{\R,r_0,R}^M(\E)$ (see Lemma~\ref{polesexist}).

Notice that $Z_{\R,r_1,R}^M(\E)$ has the maximal value
$$      d:=Z_{\R,r_1,R}^M(0)
=\R \left(1-\exp\left\{-{R\over (n+1)\R M}\right\}\right).  $$
Let $z_1$ be the first point on $\Gamma$ such that $|z_1-z_0|=d/2$ (if
no such point exists then we have finished).

Next we show that by restricting the range of $\E$ we can ensure that the
initial value problem at $z_1$ gives us a solution whose radius of
convergence in $z$ is again bounded below by $d/2\/$.
At $z=z_1\/$, ${\bf y}^{(0)}$ is analytic in $\E$ for $|\E|<\tilde r$ for some
$\tilde r<r_0$.  Let $S_1(\tilde r)
:=$ $\{(z,y_1,\ldots,y_n;\E): |z-z_0|\le\R,\ |\E|\le \tilde r,\ |y_i-
\eta_i(z_0)|\le R,\ 1\le i\le n\}$.  Then
\barray
\sup_{S_1(\tilde r)}|f_i|
&\le&   \sum    |a^i_{jk_1 \dots k_nl}| B^j\left(R+\sup_{|\E|<\tilde r}
\left|y^{(0)}_1(z_1,\E)\right|\right)^{k_1}\cdots\nn\\
& &\cdots\left(R+\sup_{|\E|<\tilde r}
\left|y^{(0)}_n(z_1,\E)\right|\right)^{k_n}r_0^l.\label{polesnewest}\earray
Now as $\tilde r\to 0$, $\sup_{|\E|<\tilde r}|y^{(0)}_i(z_1,\E)|\to
|\eta_i(z_1)|\le L$, and so \rf{polesnewest} approaches $\half M$.  Therefore
there exists $r_1$ such that $0<r_1\le\tilde r< r_0$ and
$$      \sup_{S_1(r_1)}|f_i|\le M.      $$

Invoking Lemma
\ref{polesexist} again we see that there is a solution, ${\bf y}^{(1)}$,
of equations \rf{polesdiff} satisfying
$$      {\bf y}^{(1)}(z_1,\E)={\bf y}^{(0)}(z_1,\E)$$
for all $|\E|<r_1$, which is analytic in $(z,\E)$ provided $|z-z_1|<
Z_{\R,r_1,R}^M(\E)$.  We then look for the next point, $z_2$, on $\Gamma$
such that $|z_2-z_1|=d/2$ (if such a point exists) and repeat the above
argument for a finite number of points $z_2,z_3,\ldots, z_N$ in order to
cover the curve.  ${\bf y}^{(i+1)}$ analytically continues
${\bf y}^{(i)}$.  ${\bf y}(z):=(y_1(z),\ldots,y_n(z))$ is then defined to be
${\bf y}^{(k)}$ whenever $z$ lies in the domain of analyticity of
${\bf y}^{(k)}$.  Since we proceed in steps of $d/2$ in $z$,
the radius of convergence of ${\bf y}(z)$ about any point of $\Gamma$
is bounded below.
The compactness and pathwise connectedness of $K$ then ensure that we can
analytically extend ${\bf y}(z)$ to all of $K$ by using a finite
number of curves $\Gamma_j$ from $z_0\/$.  The existence of the number
$r_K$ then follows because we require only a finite
number of reductions of $r\/$ in the above analytic continuation of
${\bf y}(z)\/$.
\begin{flushright}$\tomb$\end{flushright}
$$ $$

The condition that $K$ be simply connected is
essential for the single-valuedness of ${\bf y}(z)\/$.
For example, consider the equation
$$	y''=6y^2+\E z^2.$$
The solutions of this equation for $\E=0$ are elliptic functions and
therefore meromorphic.  However, Painlev\'e analysis (see
\cite{kruskalclarkson})
reveals that
generic solutions to this equation for
$\E\ne 0$ possess logarithmic singularities.
So locally analytic solutions of the equation with $\E=0$ whose
domain of analyticity ($\Omega$ in Theorem~1) is not simply connected
do not necessarily
arise from analytic solutions of the general equation, but rather from
multivalued ones.

In the case of $P_I$ and $P_{II}\/$, however, the
solutions are meromorphic \cite{painlevestuff, joshidirect}. 
So, on recalling the form of transformation \rf{polestrans}, we see that
all solutions of $P_{II}(\E)$ are meromorphic and therefore single
valued.  Analytically extending any solution of $P_{II}(\E)$ along any path
connecting $z_0$ to any other point will give a result which is independent of
the particular path chosen.
Hence, when we apply Theorem 1 to $P_{II}(\E)$
we can weaken the requirement that $K$ be simply connected,
instead demanding only that it be pathwise connected.  The theorem then
has the following corollary.

\begin{corollary}
Choose $z_0,\A,\B\in\complex\/$.  Let $y_I$ and $y$ be maximally extended
solutions of $P_I$ and $P_{II}(\E)$ respectively, both satisfying the initial
value problem given by
$$	y(z_0)=\A,\qquad y'(z_0)=\B.	$$
Let $\Omega\subset\complex$ be the domain of analyticity of $y_I\/$.
Given any compact $K\subset\Omega\/$, $\exists r_K>0$ such that $y$ is
analytic in $(z,\E)$ for $z\in K\/$, $|\E|<r_K$ and $y\to y_I$ with respect
to the sup norm as $\E\to 0\/$.
\end{corollary}

\proof
Apply Theorem 1 using some compact pathwise
connected subspace $\widetilde K\subset\Omega$ such that $\{z_0\}\cup
K\subseteq\widetilde K$.
\begin{flushright}$\tomb$\end{flushright}

\section{Coalescence of Poles and the Second Painlev\'e Equation}
We now return to the problem of estimating the rate of coalescence of
poles in a family of solutions to $P_{II}(\E)\/$
$$	y''=2\E^6 y^3+6y^2+\E^6zy+z,	$$
as $\E\to 0\/$.
Choose $z_0\in\complex\/$.
We will consider a family of solutions to $P_{II}(\E)$ given by $y(z_0)=\A\/$,
$y'(z_0)=\B\/$.  Multiplying $P_{II}(\E)$
through by $y'$ and integrating along some path $\C$ from $z_0$ to $z\/$
gives
\beq	\left[ y'(z)\right]^2 =
\E^6y^4+4y^3+2zy+\E^6zy^2-
{\int\limits_\C}^z\left\{2y+\E^6y^2\right\}dz+k_\E
=: \E^6F_\E\{z,y\}\label{polesFdef}\eeq
where
$$ k_\E= \B^2-\left\{\E^6\A^4+4\A^3+
2z_0\A+\E^6z_0\A^2\right\}.	$$

{}From the corollary to Theorem 1 in section 3 we see that as $\E\to 0$
the solution to $P_{II}(\E)$ given by $y(z_0)=\A\/$, $y'(z_0)=\B\/$
converges to the solution $y_I$ of $P_I$ satisfying the same initial
conditions, on any compact subset $K$ of the domain of analyticity of
$y_I\/$.

Suppose $y_I$ has a double pole at $\hat z\/$.  Let $D$ be the closed disc of
radius $\rho$ centred at $0$ containing both $z_0$ and $\hat z\/$ in its
interior.  Let $K$ be $D$ after we have deleted open discs of small radius
$\delta$ centred at each of the poles of $y_I$ which lie in $D\/$.
{}From Corollary~3 of Section~3 we see that for sufficiently small $\E\/$,
any simple pole of $y$ which lies in $D$
must be within $\D$ of a double pole of $y_I\/$ (since it cannot lie in
$K\/$).


Let $z_+, z_-\in B_\D(\hat z)$ be the positions of two poles of $y$ of
oppositely signed residues.  The distance between these poles is given by
\beq	\left|z_+-z_-\right|=\left|\int_{z_-}^{z_+}dz\right|
=\left|\E^{-3}\int_\Gamma {dy\over\sqrt{F_\E\{z,y\} }}\right|
\label{pappiieest}\eeq
for some path $\Gamma$ between points whose natural projection to ${\bf
CP}^1$ is $y=\infty\/$.  The final integral in \rf{pappiieest} makes sense
if we use the fact that {\em locally\/} on $\Gamma\/$, $z$ may be given as a 
function of $y\/$.  Also, since all solutions of $P_{II}(\E)$ are nonconstant
meromorphic functions, the points on $\Gamma$
at which $F_\E\{z,y\}$ vanishes do not accumulate.

As was the case in the coalescence of poles induced by the $E_{II}\to
E_I$ degeneracy, the opposite signs of the residues of the poles of $y$
at $z_-$ and $z_+$ indicates that $\Gamma$ must loop around a zero, $z_1\/$,
of
$y'\/$.  The close proximity of the poles for $\E$ small
indicates that $y$ must be large
at this stationary point.

We take $\Gamma$ to loop around
a zero, $z_1\/$, of $F_\E\{z,y\}\/$, to be specified
below.  Let $A:=|y(z_1)|\/$.
Define the region
$$	\Omega:=\left\{z\in D : \ 
 |y(z)| > A\right\}.	$$
Then $\Omega$
is a union of regions surrounding poles of $y\/$.
We take $z_1$ so that the connected component, $R_+\/$, of $\Omega$ whose
closure contains $z_+$ contains no other stationary points of $y\/$; {\em
i.e.}
$y'(z)\ne0$ for all $z\in R_+\/$.  An analogous argument to that outlined
in Section~2 shows that $R_+$ contains no pole other than $z_+$ and that
$z_1$ is the initial point for two curves of steepest ascent and two curves
of steepest descent (each separated by one of the four level curves of
$|y|$ which pass through $z_1\/$).  This follows from the fact that
$y(z_1)\ne0\/$, $y'(z_1)=0\/$, and $y''(z_1)\ne0\/$.
One of these steepest ascent curves,
$\Gamma_+\/$ say, lies in $R_+$ and so connects $z_1$ to $z_+\/$.  The other
steepest ascent curve, $\Gamma_-\/$, lies in $\Omega
\setminus R_+$ and is of finite
length (necessarily terminating at a pole, $z_-\/$ say)
since, for large $A\/$, $\Omega$ is a union of small disjoint
regions containing the poles of $y_I\/$.  We take the path of integration,
$\Gamma\/$, in equation \rf{pappiieest} to be the union of these two
paths.

Since $z_1$ is in the boundary of $\Omega$ and is the initial point for a
curve of steepest descent, there is a curve connecting $z_0$ to $z_1$
contained in $D\setminus\Omega\/$.  The (initial) path of integration, $\C\/$,
in equation \rf{polesFdef} connecting $z_0$ to $z\in\Gamma$ is taken to be
this descent
curve followed by one of the steepest ascent curves,
$\Gamma_1$ or $\Gamma_2\/$, from $z_1$ to the point $z\/$.


We now estimate $F_\E\{z,y\}$ for
large $y\/$ on such a curve.  For $z\in\Gamma\/$, we have
$$\sup_{\zeta\in\C}|y(\zeta)|=|y(z)|.	$$
Since $D\setminus\Omega$ only contains a finite number of small holes,
the length of the path $\C$ from $z_0$ to any point on $\Gamma$ can be
bounded by some $d>0$ which is independent of $\E$ for $\E$ small.
Also, for small $\E\/$, $k_\E$ can be bounded above by $c^2\/$, say,
where $c>0$ is
independent of $\E\/$.

{}From equation \rf{polesFdef} we see that $A:=|y(z_1)|$ is asymptotically
close to $4|\E|^{-6}\/$.
Let $r>0$ be an upper bound on $\E^6$ which is so small that
$$	r < \max \left\{d^{-1},\R^{-1}\right\},\quad\mbox{and}\quad
	A > \max \left\{c,d,\R\right\}.	$$
We now see that on $\Gamma\/$, for $|\E|^6<r\/$,
$$	\E^6F_\E\{z,y\}=\E^6y^4+4y^3+\phi(z,y),	$$
where
\barray
|\phi(z,y)| &\le & |2zy|+\left|\E^6zy^2\right|+\left|
{\int\limits_\C}^z2ydz\right|
+\left|{\int\limits_\C}^z\E^6y^2dz\right|+\left| k_\E\right|\nn\\
&\le&	2\R|y|+r\R|y|^2+2d|y|+rd|y|^2+c^2\nn\\
&\le& \kappa|y|^2,\nn
\earray
where $\kappa=5+r(d+\R)\/$.

So on $\Gamma\/$, $\phi(z,y)=\psi(z,y)y^2$ where $|\psi(z,y)|<\kappa\/$,
giving
\beq\E^6F_\E\{z,y\} = \E^6y^4+4y^3+\psi(z,y)y^2. \label{papFest}\eeq
%
%

Now $\arg(y)$
is a constant along any path of steepest ascent for $y\/$ (since
$d|y|=|dy|$ there).  Hence $\Gamma_+$
can be parameterized by $t\in (1,\infty)\/$, where $y=ty_1\/$,
$y_1:=y(z_1)\/$. 
Since $F_\E\{z_1,y_1\}=0\/$, we see from \rf{papFest} that
$$	\left|{\E^6y_1\over4}+1\right|\le{\kappa\over 4A}\to 0$$
as $\E\to0$, giving
\beq \lim_{\E\to 0}{\E^6y_1\over 4}=-1.
	\label{papfraclim}\eeq
So if we hold $t$ fixed as $\E\to0$ we have
\beq  \lim_{\E\to 0}{\E^6y\over 4}=-t.   
	\label{papfraclimtwo}\eeq
Consider the following ratio
$$\displaystyle
R_\E:={F_\E\{z,y\}\over y^3(y-y_1)}={\displaystyle
\left({\E^6y\over4}\right)^2+
\left({\E^6y\over4}\right)+{\E^6\psi\over16}\over\displaystyle
\left({\E^6y\over4}\right)^2-\left({\E^6y_1\over4}\right)
\left({\E^6y\over4}\right)}.$$
Using the limits in \rf{papfraclim} and \rf{papfraclimtwo}, we see that
$$\lim_{\E\to 0} R_\E=1.$$
The definition of limit then shows for some given $0<\nu<1\/$, 
$\exists r>0$ sufficiently small such that
for all $\E$ with $\E^6<r\/$,
$$	\left|F_\E\{z,y\}\right|\ge\nu^2\left|y^3(y-y_1)\right|.	$$
Since the same argument holds on $\Gamma_-\/$,
we have from equation \rf{pappiieest},
$$
\left|z_+-z_-\right| \le {2\over\nu|\E|^3}
\left|\int_\infty^{y_1}{dy\over\sqrt{y^3(y-y_1)}}\right|,$$
where the integration is along a path of steepest descent
(from a pole of $y\/$).  So,
recalling
that along such a path, $|dy|=-d|y|\/$, we have
$$\left|z_+-z_-\right| \le-{2\over\nu|\E|^3}
\int_\infty^A{d|y|\over\sqrt{|y|^3(|y|-A)}}={4\over\nu A|\E|^3}.	$$
Since $A\approx4|\E|^{-6}\/$,
this shows us that the distance between the poles of solutions of
$P_{II}(\E)$ is of order $\E^3\/$.

\vskip 8mm
\noindent
{\em {\bf Acknowledgements: }
The research reported in this paper was supported by the Australian Research
Council.}

%
%
%
%
\end{document}